\documentstyle[12pt,epsf]{article}
\newcommand{\be}{\begin{equation}}
\newcommand{\ee}{\end{equation}}
\newcommand{\ba}{\begin{array}}
\newcommand{\ea}{\end{array}}
\renewcommand{\P}{\mbox{\boldmath$P$}}
\newcommand{\ph}{\hat{p}}
\begin{document}
\title{Jets and Quantum Field Theory}
\author{N.A.Sveshnikov\,$^a$ and F.V.Tkachov\,$^b$
\and
{\footnotesize $^a$ Department of Physics, Moscow State University,
Moscow 119899}
\and
{\footnotesize $^b$ Institute for Nuclear Research of Russian Academy of
Sciences,
Moscow 117312}}
\maketitle
\begin{abstract}
We discuss quantum--field--theoretic interpretation of
the family of observables (the so--called C--algebra) introduced
in \cite{12} for a systematic description of multijet structure of
multiparticle  final states at high energies.  We  argue  that
from  the  point of view of general quantum field theory,  all
information about the multijet structure is contained  in  the
values  of a family of multiparticle quantum correlators  that
can be expressed in terms of the energy--momentum tensor.
\end{abstract}

\bigskip
\noindent
{\large\bf Introduction \hfill  1}
\bigskip

 The  concept  of hadron jets is a cornerstone of  the  modern
high  energy physics (see e.g. the reviews \cite{1}, \cite{2}): It would
simply  be  impossible to discuss the experimental  procedures
employed  e.g. in the recent discovery of the top  quark  \cite{3},
\cite{4}  without using the language of hadron jets. Yet apart from
the early discussion of the issue of perturbative IR safety in
connection   with   perturbative   calculability   \cite{5},   \cite{6},
remarkably  little  (if  anything at all)  has  been  done  to
integrate the jet paradigm into the framework of Quantum Field
Theory. This is despite the fact that perturbative QFT is  the
only   systematic   calculational  framework   for   obtaining
theoretical predictions about jets. The conventional theory of
jets was developed by trial and error within experimental  and
phenomenological communities and is based on the notion of jet
definition  algorithm which is foreign to QFT.  On  the  other
hand,  theoretical studies of jet algorithms and  the  related
issues  of Sudakov factorization etc. ( cf. e.g. \cite{7}, \cite{8}  and
refs.  therein ) simply accepted the conventional jet paradigm
without attempting to bring it into conformity with QFT.  This
would  be  understandable  should QFT  be  a  phenomenological
scheme  of  limited applicability. But QFT  is  a  fundamental
framework to describe mechanics of the most elementary bits of
matter  known to date that is capable of yielding  predictions
that  agree  with experiments to unprecedented precision  \cite{9},
and there are no reasons whatsoever to doubt its basic
nature.\footnote{There is, of course, gravity but the  fact  has  no
bearing on the high energy physics yet.}

 The  eclectic nature of the theory of jets did  not  fail  to
result  in practical difficulties. These are manifest in  what
is  known as ambiguities of jet definition (see e.g. \cite{1}) that
limit the precision of experimental results obtained using the
conventional data processing techniques. For instance,  it  is
expected that the dominant error of determination of  the  top
mass at the LHC will be limited by the systematic error due to
ambiguities  of  jet algorithms \cite{10}. The  difficulties  of  a
different  nature  arise  in theoretical  studies:  jet  cross
sections  are impossible to compute analytically even  in  the
simplest  cases,  while  studies  of  such  issues  as   power
corrections are unduly cumbersome (as compared with  the  non--jet case).

 It  has recently been argued (\cite{11}, \cite{12}; see also \cite{13}) that
a  systematic description of jet--related features of  hadronic
final  state in high energy physics can be achieved in a  QFT--compatible
manner within the so--called formalism of  C--algebra
(C  is  from  'calorimetric', here and below).  The  C--algebra
consists  of  a basic class of observables --- the so--called
C--correlators  that  have a rather simple form of  multiparticle
correlators with their dependence on particles' energy rigidly
fixed  (see  below)  --- plus  a few  rules  to  construct  new
observables  from  those  already  available.  The   resulting
observables possess optimal stability properties with  respect
to  data  errors  and  statistical fluctuations,  and  can  be
computed  from  events bypassing jet finding algorithms,  thus
avoiding   the  notorious  problem  of  ambiguities   of   jet
definition.  On  the other hand, the examples  given  in  \cite{12}
demonstrate  that  the expressive power of  the  C--algebra  is
sufficient  to  express practically any  jet--related  property
studied in high energy physics (such as '$n$--jet fractions'  and
mass spectra of 'multijet substates').

 As  was  pointed  out  in  \cite{12},  the  central  role  of  the
observables  of  correlator form  in  the  C--algebra  opens  a
prospect  of  a  systematic QFT study of the theory  of  jets.
However, the issue of compatibility of the C--algebra with  QFT
was  only  touched  upon in \cite{12}. There are  two  points  that
remain  to  be  clarified.  First,  we  will  show  that   all
information  about  multijet structure is  contained  in  true
quantum  correlators (the arguments of \cite{12} are incomplete  in
this  respect).  Second, we will express such  correlators  in
terms  of  the  energy momentum tensor so that  the  resulting
definition acquires a fundamental non--perturbative aspect.

\bigskip
\noindent
{\large\bf Setup \hfill  2}
\bigskip

 An  overview of the arguments that went into the construction
of  the C--algebra is given in \cite{13}. Here we only summarize the
formulas needed below.

 Let $i$ number the particles produced in an event. The event is
seen  by calorimetric detectors as a finite sequence of  pairs
 $\P=\{E_i , \ph_i\}$, where
 $E_i$ and
 $\ph_i$  are  the  $i$-th particle's energy  and  direction
(which can be represented e.g. by a point of the unit sphere).
Strictly  speaking one should bear in mind the  following  two
points.  First, such an "event" is an element  of  the  factor
space  of  the  space  of  final states  with  respect  to  an
equivalence   relation   (namely,  collinear   particles   are
calorimetrically          indistinguishable).          Second,
 $E_i$         should         be        interpreted         as
 $\left|{\vec p}_i\right|$; at high energies  one  neglects
particles'  masses and the two quantities are equivalent  from
the  point of view of their use in the description of multijet
structure.

 Following \cite{12}, we describe the C--algebra using the  language
of  particles.  Regarding  hadrons in  final  states  as  free
massive particles we have to interpret the formulas of  \cite{12}
in  the  context  of the corresponding Fock space.  The  final
formula  has  a  substantially wider meaning:  no  assumptions
about  the  structure  of the space of final  states  will  be
necessary.

 An  observable  is  defined via a  function  on  the  events,
 $f(\P)$, and the value of the observable measured for a  given
initial      state      is     obtained      by      averaging
 $f(\P)$ over the entire ensemble of events generated from that
initial state. One can always formally define (without  adding
new            information)            the            operator
 $O_f$ such that
$$
O_f\left| \P \right\rangle =f(\P)\,\left| \P
 \right\rangle
\eqno(2.1) $$
Then the measured value of the observable is
$$
\left\langle {O_f} \right\rangle \equiv \left\langle {in}
\right|O_f\left| {in} \right\rangle ,
\eqno(2.2) $$
where
$\left|  {in}  \right\rangle $ is  the  corresponding  initial
state.

 The C--algebra of \cite{12} consists of the so--called C--correlators
that  form  its  basis  and  a  few  rules  to  construct  new
observables. A C--correlator has the form
$$
F_N(\P)=\sum\nolimits_{i_1} \,\ldots \;\sum\nolimits_{i_N}
\,E_{i_1}\ldots E_{i_N}\,f_N(\ph_{i_1}\ldots \ph_{i_N})\ ,
\eqno(2.3) $$
where      $N$      is     any     positive     integer      and
$f_N(\ph_1\ldots \ph_N)$ can be any symmetric  function
of   its   $N$  angular  arguments;  without  loss  of  physical
generality        we        take        it        to        be
$C^\infty  $.  Notice  that the energy dependence  of  such  a
correlator is fixed.

 Among  the  rules to construct new elements of C--algebra  are
algebraic  combinations  as well as  integration  and  certain
forms of minimization with respect to a parameter (neither  of
which  is  of interest in the present context). A construction
of  a different type involves differential distributions which
effectively reduces to allowing new observables of the form
$$
\delta \left( {s-F(\P)} \right)\ ,
\eqno(2.4) $$
where  $s$ is a real parameter, $F$ is an observable from  the
C--algebra,  and  $\delta$  is the Dirac $\delta$--function.
(This  construction becomes  phenomenologically useful only  in
combination  with other rules but this is of no importance here).

\bigskip
\noindent
{\large\bf Expressing spectral observables
via C--correlators \hfill 3}
\bigskip

 The  first point that has to be clarified is as follows.  The
spectral observable 2.4 is, of course, to be interpreted as  a
measure,   i.e.   it  describes  a  family  of  numeric--valued
observables,  each  corresponding  to  a  continuous  function
$\chi (s)$:
$$
\int {ds\,}\chi (s)\,\delta (s-F(\P))=\chi (F(\P))
\eqno(3.1) $$
Although such an observable is indeed expressed in terms of  a
C--correlator, $F(\P)$,  but  only  at  the level of  a  single  event.
This, however,  is  not  enough  for  an  entirely  meaningful   QFT
interpretation. What one would like to be able to say is  that
the  observable after the averaging over all events  could  be
interpreted  in terms of C--correlators also taken  after  such
averaging.  The additional argument required for  this  is  as
follows.

 According to the well--known Weierstrass approximation theorem
(see   e.g.  \cite[{\it Theorem  802}]{14}),  any  continuous   function
$\chi(s)$  can  be  approximated in the  uniform  sense  by
polynomials of $s$ within a bounded interval. (Note  that  a
C--correlator is always bounded by a constant.)
So  the  observables  of the form 3.1 can be  approximated  to
arbitrary   precision   by  finite  linear   combinations   of
 $F^n(\P)\,,\;\;n\ge 0$.  If  $F$  is  an   N--correlator   then
 $F^n$ is an
 $n\times   N$--correlator.   Due   to   uniformity    of    the
approximation, one can always change the order of  taking  the
linear  combination and averaging over all events for a  given
initial state.
 The conclusion is that any differential observable from the
C--algebra  as described in \cite{12} can be regarded as appropriately
approximated  by  C--correlators after the averaging  over  all
events.  So  it becomes entirely meaningful to  say  that  all
physical  information about the multijet structure  of  events
corresponding to a given initial state is given by the average
values  of  all  C--correlators over the corresponding  events.
Therefore,  in  what follows we concentrate  on  operator  QFT
interpretation of the C--correlators 2.3.

\bigskip
\noindent
{\large\bf Operator form of C--correlators \hfill 4}
\bigskip

 One can formally rewrite 2.3 similarly to 2.2:
$$
\ba{cc}
 &\left\langle \sum\nolimits_{i_1} \,\ldots\,
\sum\nolimits_{i_N} \,E_{i_1}\ldots E_{i_N}\,f_N(\ph_{i_1}
\ldots \ph_{i_N}) \right\rangle _{\P} \ = \\
& \\
=&\int {dn_1\,}\ldots \int {dn_N\,}\left\langle {in}
\right|\varepsilon (n_1)\ldots \varepsilon (n_N)\left| {in}
\right\rangle \times f_N(n_1,\ldots ,n_N)\ ,
\ea
\eqno(4.1) $$
where  all  $n$  are  unit 3--vectors (points  of  unit  sphere),
$\varepsilon  (n)$ is an operator--valued distribution  on  the
unit    sphere.    It    is    a   QFT    interpretation    of
$\varepsilon (n)$ that we wish to obtain.

 In the context of the asymptotic Fock space, one has
$$
\varepsilon (n)=\int {{{dp} \over {2p^0}}}\,a^+(p\,)\;a^-
(p\,)\times \left( {\ph\cdot n} \right)\;\delta (\ph,n) ,
\eqno(4.2) $$
where  the  last factor is the $\delta$--function on the  unit  sphere
localized              at              the               point
$\ph=n$.  One notices that in presence of such $\delta$--function,
$\left(  {\ph\cdot n} \right)=\left| {\,{\vec p}\,} \right|\approx~E$
(recall   that   any  definition  of  jets   and   jet--related
observables is valid only in the limit of high energies  where
all   particles  are  regarded  as  massless).  The  operators
$a^\pm  $  are  to  be  interpreted  as  free--field  operators
corresponding     to     the     asymptotic     states      at
$t=+\infty $.

\newpage
\noindent
{\large\bf {\boldmath$\varepsilon (n)$} in terms of fields.
A heuristic derivation. \hfill 5}
\bigskip

 Without  loss  of  generality we take  all  particles  to  be
identical  Lorentz--scalar bosons. We also limit  ourselves  in
this  first  heuristic derivation to particles  with  non--zero
mass   $m$.   Consider   the  quantum  field   corresponding   to
$a^\pm $:
$$
\varphi (x,t)=\int {{d^3 k} \over {2 k^0}}\,\left(
{e^{+ikx}a^+(k)+H.C.} \right).
\eqno(5.1) $$
(We    dropped   irrelevant   normalization   factors.)   Take
$x={p  \over  p^0}t\,,\;\;t\to  +\infty  $,  and
formally use the stationary phase approximation. Then:
$$
\varphi \left( {p \over {p_0}}t,\;t
\right)\mathop \approx \limits_{t\to +\infty }{{p_0^{3/2}}
\over {t^{3/2}}}\times {1 \over {2m}}\times \left(
{a^+(p)e^{i{{m^2t} \over {p_0}}+
i{{3\pi} \over 4}}+H.C.} \right)
\eqno(5.2) $$
The combination
$a^+(p\,)\;a^-(p\,)$ can be extracted as follows:
$$
\partial _0\varphi \,\partial _i\varphi \left(
{p \over {p_0}}t,\;t \right)=t^{-3}\left\{
{{{p_ip_0^4} \over {2m^2}}a^+(p)\,a^-(p)
+O\left( {e^{\pm
2i{\textstyle{{m^2t} \over {p_0}}}}} \right)} \right\}+o\left(
{t^{-3}} \right)
\eqno(5.3) $$
Noticing  that  the  combination of field derivatives  on  the
l.h.s.                        is                       exactly
$T_{0i}$,  the  energy--momentum  tensor  of  the  free  field
$\varphi $, and formally neglecting the oscillating terms, one
obtains:
$$
\varepsilon (n)=m^2t^3\int {{{dp} \over {p_0^5}}\,}\delta
(\ph,n)\,T_{0i}\left( {{\textstyle{p \over {p_0}}}t,\;t}
\right).
\eqno(5.4) $$
The   power   of   mass   can  be  eliminated   by   rescaling
$p\to mp$.
 There are two subtle points that have to be clarified in  the
above derivation:
\begin{itemize}
\item  The derivation has to be valid for massless field, and for
$m=0$ one should be able to take into account the rather
complex singularity of the field near the light cone, namely,
$$
\varphi (x,t)\sim
\left\{
\ba{l}
t^{-1}\,,\;\,|x\,|=t\,,\\
t^{-2}\,,\;\,|x\,|<t\,.
\ea
\right.
\eqno(5.5) $$
\item How  to  accurately neglect the oscillating terms  on  the
r.h.s. of 5.3?
\end{itemize}
 These are the two issues that we are now going to consider.

\bigskip
\noindent
{\large\bf A more accurate derivation \hfill 6}
\bigskip

 Motivated  by  the above derivation, consider  the  following
expression (a smearing over $n$ is implicit):
$$
\mathop {\lim }\limits_{t\to \infty }t^3\int\limits_0^1
{\rho ^2d\rho \,}\partial _0\varphi \partial _i\varphi \left(
{n\rho t,t} \right)
\eqno(6.1) $$
Substituting 5.1 one obtains the following expression for  the
coefficient                                                 of
$a^\pm (p)\,a^\pm (q)$:
$$
\int\limits_0^1 {\rho ^2d\rho \,}\int {{{d^3 p} \over
{2p_0}}\,{{d^3 q} \over {2q_0}}\,\,e^{it\left[ {\pm \left( {p_0-
pn\rho } \right)\pm \left( {q_0-qn\rho } \right)}
\right]}}\;p_0\,q_i\,a^\pm (p)a^\pm (q).
\eqno(6.2) $$
The      asymptotics      of     this      expression      for
$t\to  +\infty  $  can  be found using  the  stationary  phase
method.

 One  finds that the stationary points are as follows.\\
For the term
$a^\pm  (p)\,a^\pm  (q)$ with opposite signs  the  stationary
point                                                       is
$p=q={\textstyle{{m\rho } \over {\sqrt {1-\rho ^2}}}}n$  with
any $\rho\in [0,1]$ for $m\ne 0$, and $\rho=1$,
$p=q=\omega n$  with              any
$\omega >0$                        for
$m=0$.               For               the               term
$a^\pm (p)\,a^\pm (q)$ with both signs equal one should  take
$\rho                         =0$                         and
$p=q=0$                       for                        both
$m\ne 0$ and $m=0$,  and the entire contribution is suppressed as compared
with   the   previous   case  by  an   additional   power   of
$t^{-1}$. This settles the problem of oscillating terms.
 Finally, one arrives at the following expression irrespective
of the field's mass:
$$
\varepsilon (n)\,dn\,=\mathop {\lim }\limits_{t\to +\infty
}\int\limits_0^t {\rho ^2d\rho \,}n_iT_{0i}\left( {\rho n,t}
\right)dn\,=\mathop {\lim }\limits_{t\to +\infty
}\int\limits_{x\in Cone(t,n,dn)} {dx\,}n_iT_{0i}\left( {x,t}
\right)\,dn
\eqno(6.3) $$
To  help  understand  this expression, below  is  a  graphical
representation       of      the      integration       region
$x\in Cone(t,n,dn)$ on the r.h.s.:
\begin{figure}[t]
\epsfverbosetrue
\epsfysize=8cm
\epsfbox{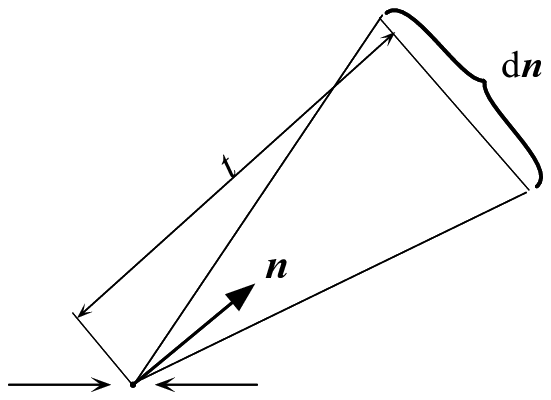}
\end{figure}
Recall                                                    that
$T_{0i}(x)$  is  the space--time density of 3--momentum  and  is
equal  due to symmetry of the energy--momentum tensor    to
$T_{i0}(x)$,  the  density  of energy  flow.  The  convolution
$n_iT_{0i}$  projects out the component of  the  flow  in  the
direction  $n$.  The  integration has to be performed  over  the
entire  cone  including small $x$ (which is the  least  expected
feature  of  the  answer).  This  is  apparently  because,  in
general, particles in the final state may be arbitrarily slow.

\bigskip
\noindent
{\large\bf Conclusions \hfill  7}
\bigskip

 We  have  demonstrated  that all physical  information  about
multijet  structure of multiparticle events at  high  energies
represented by the C--algebra of observables described in  \cite{12}
is  contained in the family of the so-called C--correlators  of
the                form               4.1                where
$f_N$  are  arbitrary  smooth  symmetric  functions  of  their
angular        arguments,       and        the        operator
$\varepsilon (n)$ is defined by 6.3. Such a definition of
C--correlators  retains physical meaning even if the theory  does
not  admit a naive particle interpretation (e.g. QCD where the
asymptotic  states cannot contain quark and gluon fields;  cf.
e.g.  the  discussion in \cite{15}; or QED where the  structure  of
asymptotic states is rather complex due to massless  photons).
The  definition of jet--related observables in terms of the
C--correlators  4.1,  6.3  is  non--perturbative  which  fills  an
important gap that remained in the theory of QCD jets.

 To reiterate, we started with the assumption that hadrons can
be  regarded as free particles in final states. Then we showed
that  that  the C--correlators are expressed in  terms  of  the
energy  momentum  tensor of the corresponding asymptotic  free
fields via formulas that are universal in the sense that  they
are  independent of particles' masses. Therefore, summing over
all  particles' types yields the total energy momentum tensor.
The   latter,   however,  being  related  to  the   space--time
symmetries of the theory, is independent of the set of  fields
used  to formulate the theory. In particular, one can use  its
expression in terms of quark and gluon fields as well.

 All this puts the issue of hadronization effects in jet cross
sections into a clearer perspective. For instance, in the case
of
$e^+e^-$  annihilation into hadrons the bound states  do  not
appear  in  the  expressions 4.1, 6.3 at  all.  Likewise,  the
important \cite{16} issue of the structure of power corrections can
be  systematically  studied  (limiting  the  consideration  by
necessity  to  perturbation theory) following the  pattern  of
\cite{17}  within the systematic formalism of asymptotic  operation
\cite{18}   appropriately  modified  for  non--Euclidean  asymptotic
regimes  \cite{19,20}. In particular, the form of 6.3  (cf.  the
integration over a half axis stretching from zero to infinity)
seems to add plausibility to the recent hypothesis that string
operators should play a role in power corrections to jet cross
sections \cite{21}.

 Lastly,  it  would  be  interesting to find  out  under  what
general assumptions one can prove existence of objects like
C--correlators 4.1, 6.3 in the context of axiomatic QFT.  It  may
be  expected that this can be done under assumptions  somewhat
weaker  than the usual one about a non-zero mass  gap  in  the
standard scattering theory (cf. \cite{22}).

\bigskip
\noindent
{\large\bf Acknowledgments}
\bigskip

This work was made possible in parts by the
International Science Foundation (grants MP9000/9300) and  the
Russian Fund for Basic Research (grant 95-02-05794a).

References


\begin{thebibliography}{66}
\bibitem{1} {\it S.D.Ellis}: talk at "QCD and High Energy Hadronic
Interactions. The XXVIIIth
   Rencontres de Moriond" (also preprint CERN-TH.6861/93).
\bibitem{2} {\it R.Barlow}: Rep. Prog. Phys. {\bf 36} (1993) 1067.
\bibitem{3} {\it CDF Collaboration (F.Abe et al.)}: Phys. Rev. Lett. {\bf 74}
   (1995) 2626.
\bibitem{4} {\it D0 Collaboration (S.Abachi et al.)}: Phys. Rev. Lett. {\bf 74}
   (1995) 2632.
\bibitem{5} {\it G.Sterman and S.Weinberg}: Phys. Rev. Lett. {\bf 39} (1977)
   1436.
\bibitem{6} {\it G.Sterman}: Phys. Rev. {\bf D17} (1978) 2789;
{\bf D17} (1978) 2773;   {\bf D19} (1979) 3135.
\bibitem{7} {\it S.Catani, Yu.L.Dokshitzer, M.Olsson, G.Turnock, and
   B.R.Webber:} Phys. Lett. {\bf 269B} (1991) 432.
\bibitem{8} {\it S.Bethke, Z.Kunszt, D.E.Soper, and W.J.Stirling:}
Nucl. Phys. {\bf B370} (1992) 310.
\bibitem{9} {\it T.Kinoshita and W.B.Lindquist}: Phys. Rev. {\bf D27} (1983)
   853.
\bibitem{10} {\it F.Dydak}: talk at the IX Int. Workshop on High
   Energy Physics (INP MSU), Zvenigorod, Russia, 16-22 Sept.
   1994.
\bibitem{11} {\it F.V.Tkachov}: Phys. Rev. Lett. {\bf 73} (1994) 2405.
\bibitem{12} {\it F.V.Tkachov}: preprint FERMILAB-PUB-95/191-T.
\bibitem{13} {\it F.V.Tkachov}: talk at the X Int. Workshop on High
   Energy Physics (INP MSU), Zvenigorod, Russia, 20-26 Sept.
   1995.
\bibitem{14} {\it L. Schwartz}: Analyse Mathematique. Vol. 1. Paris:
   Hermann, 1967.
\bibitem{15} {\it A.Yu.Kamenschik and N.A.Sveshnikov}: Phys. Lett.
   {\bf 123B} (1983) 255.
\bibitem{16} {\it B.R.Webber, in}: New Techniques for Calculating
   Higher Order QCD Corrections (ed. Z.Kunszt). ETH: Z\"urich,
   1992.
\bibitem{17} {\it F.V.Tkachov}: Phys. Lett. {\bf 125B} (1983) 85.
\bibitem{18} {\it F.V.Tkachov}: Int. J. Mod. Phys. {\bf A8} (1993) 2047;
   {\it G.Pivovarov and F.V.Tkachov}: Int. J. Mod. Phys. {\bf A8} (1993)
   2241.
\bibitem{19} {\it J.C.Collins and F.V.Tkachov}: Phys. Lett. {\bf 294B}
   (1992) 403.
\bibitem{20} {\it F.V.Tkachov, in}: New Techniques for Calculating
   Higher Order QCD Corrections (ed. Z. Kunszt). ETH: Z\"urich,
   1992; Sov. J. Nucl. Phys. {\bf 56} (1993) 180; Sov. J. Part.
   Nuclei {\bf 25} (1994) 649.
\bibitem{21} {\it G.P.Korchemsky and G.Sterman}: Nucl. Phys. {\bf B437}
   (1995) 415.
\bibitem{22} {\it D.Iagolnitzer}: Scattering in Quantum Field
   Theories. Princeton Univ. Press, 1993.
\end{thebibliography}
\end{document}